\documentclass{siamltex}

\usepackage{amsmath, latexsym, amsfonts, amssymb, amscd,bbm}
\usepackage[english]{babel}
\usepackage[dvips]{graphicx}
\usepackage[mathlines]{lineno}

\usepackage{amsmath,amsfonts,bm,bbm}

\usepackage{float}
\usepackage{epsfig}
\usepackage{esint}
\usepackage{color}

\usepackage{multirow}

  \usepackage{paralist}
  \usepackage{epstopdf}
  \usepackage{graphics} 
 \usepackage[colorlinks=true]{hyperref}
 \hypersetup{urlcolor=blue, citecolor=red}
 \usepackage[latin1]{inputenc}
\usepackage[caption=false]{subfig}

 \usepackage{tikz-cd}

\usepackage{float}
\usepackage{epsfig}
\usepackage{esint}

\usepackage{bm,braket}

\newcommand{\R}{{\mathbb R}}

\newcommand{\e}{{\rm e}}

\newcommand{\bary}{\overline{y}}

\renewcommand{\P}{\mathbb{P}}

\renewcommand{\theequation}{\arabic{section}.\arabic{equation}}

\newcommand{\E}{{\mathbb E}}

\title{Slow-fast systems with stochastic resetting}
\author{Paul C. Bressloff\thanks{Department of Mathematics, Imperial College London, London SW7 2AZ, UK  ({\tt p.bressloff@imperial.ac.uk}).}}

\begin{document}

\maketitle

\begin{abstract} 
In this paper we explore the effects of instantaneous stochastic resetting on a planar slow-fast dynamical system of the form $\dot{x}=f(x)-y$ and $\dot{y}=\epsilon (x-y)$ with $0<\epsilon \ll 1$. We assume that only the fast variable $x(t)$ resets to its initial state $x_0$ at a random sequence of times generated from a Poisson process of rate $r$. Fixing the slow variable, we determine the parameterized probability density $p(x,t|y)$, which is the solution to a modified Liouville equation. We then show how for $r\gg \epsilon$  the slow dynamics can be approximated by the averaged equation $dy/d\tau=\E[x|y]-y$ where $\tau=\epsilon t$, $\E[x|y]=\int x p^*(x|y)dx$ and $p^*(x|y)=\lim_{t\rightarrow \infty}p(x,t|y)$. We illustrate the theory for $f(x)$ given by the cubic function of the FitzHugh-Nagumo equation. We find that the slow variable typically converges to an $r$-dependent fixed point $y^*$ that is a solution of the equation $y^*=\E[x|y^*]$. Finally, we  numerically explore deviations from averaging theory when $r=O(\epsilon)$. \end{abstract}
\maketitle


\section{Introduction}

A topic of increasing interest within the theory of nonequilibrium systems is the dynamics of stochastic processes under instantaneous Poissonian
resetting. One of the simplest example of such a process is a Brownian particle whose position is reset to its initial position at a random sequence of times generated by a Poisson process with constant rate $r$ \cite{Evans11a,Evans11b,Evans14}. More specifically, let $X(t)\in \R$ denote the position of the Brownian particle at time $t$. Between resetting events, the position evolves according the stochastic differential equation (SDE)
$dX(t)=\sqrt{2D}dW(t)$, 
where $W(t)$ is a Wiener process and $D$ is the diffusivity. The SDE is supplemented by the reset condition $X(T_n^-) \rightarrow X(T_n)=x_0$ at the sequence of times $T_n$, $n\geq 1$ with $T_n-T_{n-1}$ an independent exponentially distributed random variable. One major consequence of resetting is that the corresponding probability density $p(x,t)$, which satisfies a modified diffusion equation, converges to a nonequilibrium stationary state (NESS) that supports nonzero probability currents. Moreover, the approach to the stationary state exhibits a dynamical phase transition, which takes the form of a traveling front that separates spatial regions for which the probability density has relaxed to the NESS from those where it has not \cite{Majumdar15}. The existence of a nontrivial NESS has subsequently been shown for a wide range of stochastic processes with resetting, see the review \cite{Evans20}. Examples include non-diffusive processes such as Levy flights \cite{Kus14} and run-and-tumble processes \cite{Evans18,Bressloff20}, switching diffusions \cite{Bressloff20a}, non-Poissonian resetting protocols \cite{Eule16,Pal16,Nagar16}, and diffusion in a potential \cite{Pal15}. Note, however, that most of these studies are restricted to one-dimensional systems.

Another important application of resetting arises within the context of integrate-and-fire (IF) models of neurons \cite{Keener81,Burkitt06}. In contrast to the models considered above, resetting to an initial state occurs whenever a state variable (membrane voltage) crosses some fixed threshold $\kappa$ from below -- a deterministic resetting protocol. In the case of a simple leaky IF model, we have $\dot{x}=I_0-\gamma x$, where $I_0$ is a fixed external input and $\gamma$ is a decay rate, such that $x(t)=0$ whenever $x(t^-)=\kappa$. If $0<\kappa <I_0/\gamma$, then this generates an oscillator with period $\tau=\gamma^{-1}\ln [I_0/(I_0-\kappa)]$ \cite{Keener81,Burkitt06}. Stochastic versions of IF models typically consider the effects of fluctuating inputs or thresholds on the reset times, rather than imposing an explicit stochastic resetting rule. The latter would result in a piecewise deterministic dynamical system with stochastic resetting. Such a class of model has recently arisen within the context of noiseless Kuramoto phase oscillators subject to global stochastic resetting \cite{Sarkar22,Bressloff24a,Bressloff25}. That is, all the phases are simultaneously reset to their initial values at a random sequence of times generated from a common Poisson process. (Global resetting is distinct from local resetting where the oscillators  are independently reset \cite{Bressloff24a,Ozawa24}. Another scenario is to reset only a subpopulation of oscillators \cite{Majumder24}.) In the mean field limit, the dynamics can be projected onto a low-dimensional Ott-Antonsen (OA) manifold \cite{Ott08}, in which the macroscopic state evolves according to a nonlinear piecewise-deterministic dynamical system with resetting. One thus obtains a phase diagram for the macroscopic state that is based on the structure of the corresponding NESS -- the latter is the steady-state solution of a modified Liouville equation. In the case of the classical Kuramoto model with sinusoidal coupling \cite{Kuramoto84,Strogatz00} and global resetting, the OA dynamics is effectively one-dimensional \cite{Sarkar22,Bressloff24a}. On the other hand, when the oscillators are indirectly coupled via a common external medium, the resulting OA dynamics is at least two-dimensional \cite{Schwab12,Bressloff25}. Moreover, if the spatial distribution of oscillators across the medium is sufficiently sparse, the OA dynamics reduces to a slow-fast dynamical system in which only the fast variable is subject to resetting \cite{Bressloff25}. This reflects the fact that the external medium does not reset.

The last example motivates the current study, namely, analyzing slow-fast dynamical systems with fast resetting. In order to develop the basic theory, we assume that the underlying deterministic system is of the nondimensionalized form $\dot{x}=f(x)-y$ and $\dot{y} =\epsilon (x-y)$ with $0<\epsilon \ll 1$. The deterministic dynamics is supplemented by the stochastic resetting rule  $(x(T_n^-),y(T_n^-))\rightarrow (x_0,y(T_n))$ at the Poisson generated sequence of times $T_n$. For the sake of illustration, we take $f(x)$ to be the cubic $f(x)=x(a-x)(x-1)+I_0$ for constants $a,I_0$ with $0<a<1$. The latter yields the FitzHugh-Nagumo (FN) equation that is often used to model action potential generation in neurons \cite{Terman}. Using a separation of time-scales we obtain the following results for $\epsilon \ll 1$:
\medskip

\noindent 1. The fast dynamics reduces to a 1D dynamical system with resetting in which $y=\bary $ is fixed. Rather than rapidly converging to a stable branch of the nullcline $\bary=f(x)$, the fast variable $x$ is randomly distributed according to a probability density $p(x,t|\bary)$, -- the latter satisfies a modified Liouville equation. The support of $p(x,t|\bary)$ is from the reset point $x_0$ to a root of $\bary=f(x)$.
\medskip

\noindent 2. An explicit expression for the the NESS $p^*(x|\bary)=\lim_{t\rightarrow \infty}p(x,t|\bary)$ can be obtained for polynomial $f(x)$ by factorizing $f(x)-\bary$ and using partial fractions. In addition, the relaxation to the NESS as a function of $x$ can be determined using the so-called accumulation time for dynamical systems with stochastic resetting \cite{Bressloff21}.
\medskip

\noindent 3. If the resetting rate $r\gg \epsilon$, then the slow dynamics can be approximated by the averaged equation $dy/d\tau=\E[x|y(\tau)]-y(\tau)$, where $\tau=\epsilon t$ and $\E[x|y]=\int_{\R}xp^*(x|y)dx$. The slow dynamics may induce a fast switch in the support of $p^*(x|y)$. Moreover, $y(\tau)$ typically converges to an $r$-dependent fixed point $y^*$ that is given by a root of the equation $y^*=\E[x|y^*]$.
\medskip

\noindent 4. The averaged equation no longer holds when $r=O(\epsilon)$. If the underlying deterministic system has a unique stable fixed point, then each reset triggers a trajectory that converges from $x_0$ towards the fixed point. In the case of the FN equation operating in an excitable regime this may result in a sequence of reset-triggered action potentials. On the other hand, when the FN equation operates in an oscillatory regime, resetting induces stochastic fluctuations about the underlying stable limit cycle.

\setcounter{equation}{0}

\newpage
\section{The basic setup}
Consider a slow-fast dynamical system of the form
\begin{align}
\label{fs}
 \frac{dx}{dt}&=f(x,y),\qquad \frac{dy}{dt}=\epsilon g(x,y),
\end{align}
with $0<\epsilon \ll 1$ and $x,y\in \R$. Here $x$ is the fast variable and $y$ is the slow variable. We assume throughout that time is non-dimensionalized. A classical method for studying such systems is {\em geometric singular perturbation theory} \cite{Fenichel79,Jones95,Wech20}. This involves combining the analysis of equations (\ref{fs}) in the singular limit $\epsilon \rightarrow 0$ with the corresponding analysis of the equivalent dynamical system
\begin{align}
\label{sf}
\epsilon \frac{dx}{d\tau}&=f(x,y),\qquad \frac{dy}{d\tau}= g(x,y)
\end{align}
where $\tau= \epsilon t $ is a slow time variable. Equations (\ref{fs}) and (\ref{sf}) are known as the fast and slow systems, respectively. The singular limit of equation (\ref{fs}) yields the 1D layer problem
\begin{align}
\label{fs0}
\frac{dx}{dt}&=f(x,y),\qquad\frac{dy}{dt}=0,
\end{align}
in which the slow variable $y$ is treated as a parameter. For fixed $y=\overline{y}$ one determines the fixed points of the 1D system
and their stability, exploring the onset of bifurcations as the parameter $\overline{y}$ is varied. 
 The singular limit of equations (\ref{sf}) results in the 1D reduced problem
\begin{align}
\label{sf0}
0&=f(x,y),\qquad \frac{dy}{d\tau}=g(x,y),
\end{align}
The fast process is assumed to occur so quickly that it adjusts instantaneously to changes in the slow process. We thus have a 1D dynamical system for the slow process $y$ and an algebraic equation that constrains the fast process to lie on the nullcline (slow manifold) $f(x,y)=0$. Hence, the reduced system describes the slow evolution of the system as it moves along the nullcline.

The geometrical aspect of the analysis refers to the fact that it is necessary to take into account the detailed structure of the slow manifold in order to match the reduced and layer solutions. In the 1D case this typically involves the critical points of the $f(x,y)$ nullcline with respect to the fast variable $x$. That is, when the layer solution reaches an extremum of the slow manifold it has nowhere else to go, which marks the departure point into the fast system (\ref{fs}). The subsequent motion involves
the rapid evolution of $x$ on horizontal lines that terminate at points of intersection
with the slow manifold.

\begin{figure}[t!]
\centering
\includegraphics[width=8cm]{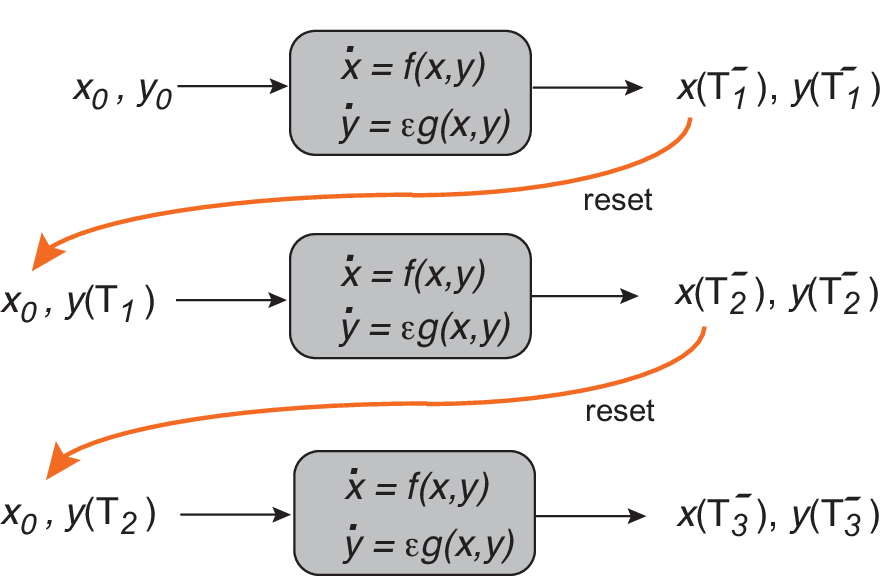} 
\caption{Schematic diagram of a slow-fast system with stochastic resetting of the fast variable. Only the first few resetting events are shown.} 
\label{fig1}
\end{figure}

In this paper we consider how the slow-fast dynamics outlined above is modified under a stochastic resetting protocol that is only applied to the fast variable. That is, the variable $x(t)$ instantaneously returns to its initial value $x_0$ at a random sequence of times generated from a Poisson process with constant rate $r$. More precisely, 
let $T_{n}$ denote the $n$th resetting time with $n\geq 1$ and $T_0=0$. The inter-reset times $ \Delta_n ={T}_n- {T}_{n-1}$ are exponentially distributed with
\begin{equation}
 \P[ {\Delta}_n\in [s,s+ds]]=r\e^{-r s}ds
\end{equation} 
 In addition, the number of resets occurring in the time interval $[0,t]$ is given  by the Poisson process 
 \begin{align}
 {N}(t)=n,\quad  {T}_{n}\leq t < {T}_{n+1},\quad \P[ {N}(t)=n]=  \frac{(r t)^{n}\e^{-r t}}{n!},
\label{Pois}
\end{align}
with $\E[N(t)]=r t=\mbox{Var}[N(t)]$. Note that $N(t)$ is defined to be right-continuous: $N (T_n^-)=n-1$ whereas $N(T_n)=n$.
Instantaneous resetting can be introduced into the fast system (\ref{fs}) by defining the differential
\begin{equation}
dN(t)=h(t)dt,\quad h(t)=\sum_{n\geq 1}\delta(t-T_n),
\label{hN}
\end{equation}
and taking\begin{align}
\label{fsresetgen}
 \frac{dx}{dt}&=f(x,y)+h(t)[x_0-x(t^-)],\qquad \frac{dy}{dt}=\epsilon g(x,y),
\end{align}
A schematic illustration of the stochastic resetting protocol is shown in Fig. \ref{fig1}. Since only the fast variable resets we have an example of subsystem stochastic resetting. Finally, in order to develop the basic theory we focus on the particular class of slow-fast equations
\begin{align}
\label{fsreset}
 \frac{dx}{dt}&=f(x)-y+h(t)[x_0-x(t^-)],\qquad \frac{dy}{dt}=\epsilon(x-y)
\end{align}
where $f(x)$ is a finite-order polynomial in $x$. We also assume that $r$ is independent of $\epsilon$ so that resetting can be treated as part of the fast dynamics in the limit $\epsilon \rightarrow 0^+$.

\subsection{The layer problem with resetting}

 In the singular limit $\epsilon \rightarrow 0$ we fix the slow variable $y(t)=\bary$, which leads to a 1D layer problem with resetting:
\begin{align}
\frac{dx}{dt}&=f(x)-\bary+h(t)[x_0-x(t^-)],
\label{xreset}
\end{align}
where $\bary$ is treated as a parameter. In order to check that equation (\ref{xreset}) implements instantaneous resetting,  we integrate equation (\ref{xreset}) with respect to $t$:
\begin{align}
 x(t) &=x_0+\int_0^t  f(x(s),\bary) ds +\sum_{m=1}^{N(t)}(x_0-x(T_{m}^-)).
 \end{align}
Setting $t= {T}_n$ and $t= {T}_n^-$, respectively, and subtracting the resulting pair of equations shows that $x( {T}_n)-x( {T}_{n}^-)=x_0 - x( {T}_{n}^-)$, that is, $x( {T}_n)=x_0$. A crucial element in establishing this equivalence is the right-continuity of the Poisson process $N(t)$.
Let $p(x,t)dx=\P[x<x(t)<x+dx]$. It can be shown that $p$ evolves according to
the generalized Liouville equation (see appendix A)
\begin{equation}
\frac{\partial p(x,t)}{\partial t}=-\frac{\partial [(f(x)-\bary)p(x,t)]}{\partial x}-rp(x,t)+r \delta(x-x_0).
\label{Liou}
\end{equation}
with $p(x,0)=\delta(x-x_0)$. (Note that $p(x,t)$ is parameterized by $\bary$. When we wish to make this explicit we write $p=p(x,t|\bary)$.)

 \begin{figure}[b!]
\centering
\includegraphics[width=12cm]{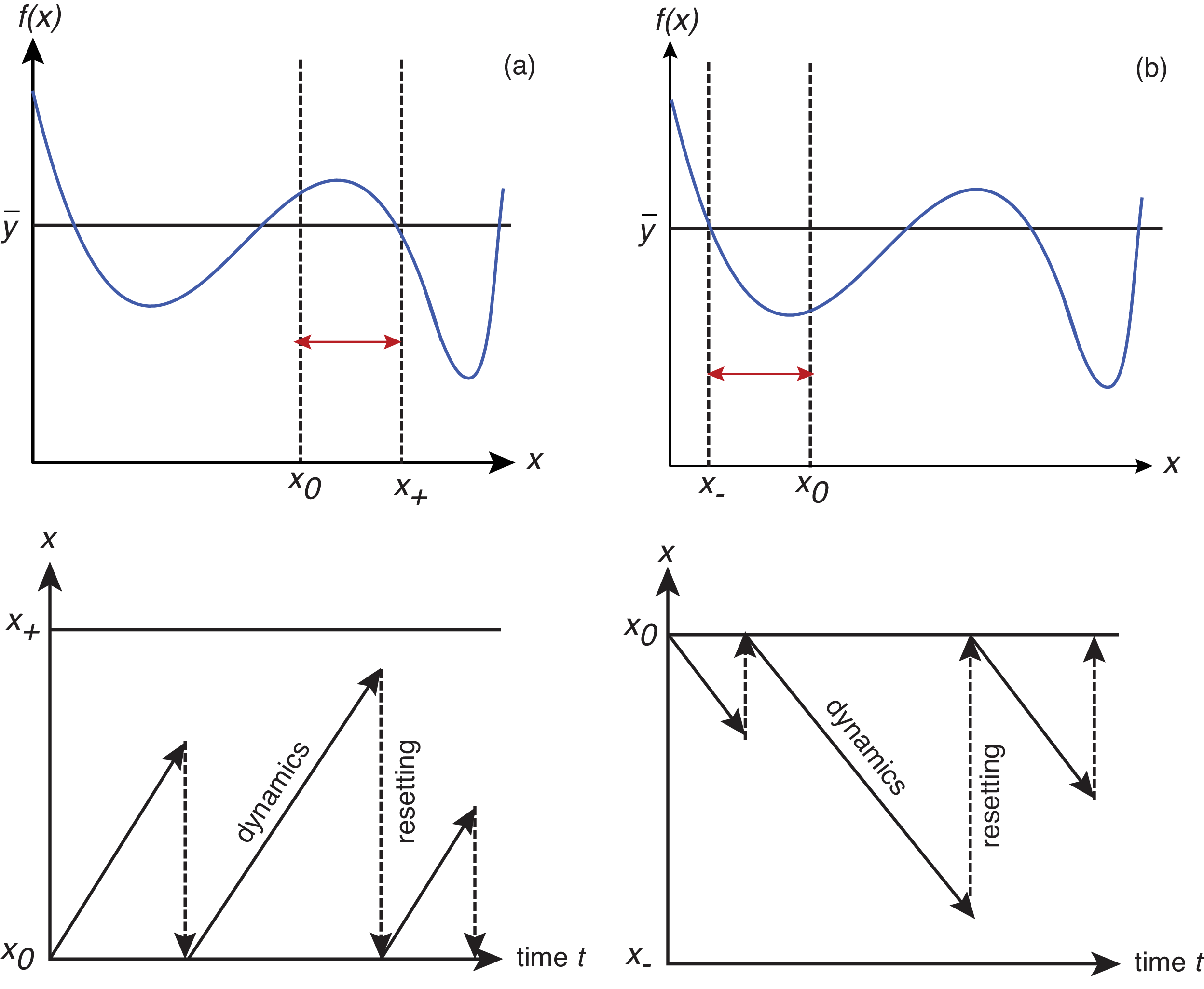} 
\caption{Schematic diagram illustrating the layer dynamics with resetting for an arbitrary function $f(x)$ and fixed $\bary$. (The diagonal arrows indicate the points at which resetting occurs -- they do not represent actual paths between resetting events.) (a) The dynamics is restricted to the interval $[x_0,x_+)$. (b) The dynamics is restricted to the interval $(x_-,x_0]$. }
\label{fig2}
\end{figure}

Assuming that resetting occurs at a rate $r=O(1)$, it follows that in the limit $\epsilon \rightarrow 0^+$, many resetting events occur while $y(t)$ hardly changes from the current value $\bary$. This suggests that the slow dynamics ``sees'' the corresponding nonequilibrium stationary state (NESS) $p^*(x)=\lim_{t\rightarrow\infty}p(x,t)$ with
\begin{equation}
\frac{d[f(x)-\bary]p^*(x)}{d x}+r p^*(x)=r \delta(x-x_0).
\label{NESS}
\end{equation}
Setting $q(x)=[f(x)-\bary]p^*(x)$, we have
\begin{equation}
\frac{dq(x)}{d x}+ \frac{rq(x)}{f(x)-\bary}=r \delta(x-x_0),
\label{qr}
\end{equation}
Solving equation (\ref{qr}) on either side of $x_0$ using an integrating factor gives
\begin{align}
p^*(x)=\left \{ \begin{array}{cc}  \frac{\displaystyle  A_-\Gamma(x)}{\displaystyle f(x)-\bary}  &  \quad x<x_0 \\ & \\   \frac{\displaystyle A_+ \Gamma(x)}{\displaystyle f(x)-\bary} & x>x_0\end{array}
\right . ,
\label{pF}
\end{align}
where
\begin{align}
\Gamma(x):&=\exp\left (-r\int^x \frac{ds}{f(s)-\bary}\right ).
\label{Fir}
\end{align}
Second, integrating equation (\ref{qr}) over the interval $ [x_0-\epsilon, x_0+\epsilon]$ and taking the limit $\epsilon \rightarrow 0^+$
implies that
\begin{equation}
A_+ -A_-=\frac{r }{\Gamma(x_0)}.
\end{equation}
This equation together with the normalization condition
\begin{equation}
\int_{\R}p^*(x)dx=1
\label{norm}
\end{equation}
determine the unknown coefficients $A_{\pm}$. 

One feature missing from the above formulation is specifying the support of $p(x,t|\bary)$ and hence $p^*(x|\bary)$. This plays a crucial role in our subsequent analysis. It turns out that the support depends on the location of the point $(x_0,\bary)$ with respect to the various branches of the nullcline $y=f(x)$. This is illustrated in Fig. \ref{fig2} for an arbitrary function $f(x)$. The roots of the equation $f(x)=\bary$ correspond to the points of intersection between the curves $y=f(x)$ and $y=\bary$. Each root $x^*$ represents a fixed point of the layer problem without resetting, which is stable (unstable) when $f'(x^*)<0$ ($f'(x^*)>0$). Suppose that $x_0$ lies in the basin of attraction of a stable fixed point $x_+$ with $x_0<x_+$. Each solution of the layer problem with resetting will alternate between a deterministic trajectory converging towards $x_+$ and an instantaneous reset to $x_0$. Hence, the stochastic dynamics is restricted to the interval $[x_0,x_+)$, which determines the support of $p^*(x,\bary)$. It follows that $A_-=0$ and the Dirac delta function on the right-hand side of equation (\ref{NESS}) represents a constant flux injected at the end $x=x_0$. The NESS becomes
\begin{equation}
p^*(x)= \frac{\displaystyle r\Gamma(x)}{\displaystyle [f(x)-\bary]\Gamma_0(x)}=-\frac{\Gamma'(x)}{\Gamma(x_0)},\quad x\in [x_0,x_+).
\label{pstar0}
\end{equation}
Similarly, if $x_0$ lies in the basin of attraction of a stable fixed point $x_-$ with $x_-<x_0$, then the support of $p(x,t|y)$ is $(x_-,x_0]$ and $A_+=0$. These two cases are illustrated in Figs. \ref{fig2}(a,b), respectively. In scenarios where the right-most fixed point $x_R$ is unstable (not shown in Fig. \ref{fig2}) and $x_0>x_R$ then some trajectories may shoot off to $\infty$ and the support of the probability density is $[x_0,\infty)$. Similarly, if the left-most fixed point $x_L$ is unstable and $x_0<x_L$ then the support is $(-\infty,x_0]$. In both of these cases, it is necessary to modify the normalization condition, since the measure at infinity may be non-zero.

\subsection{Reduced problem with resetting} The above formulation of the layered problem with resetting implies that we can no longer take the slow dynamics to be located along the nullcline $y=f(x)$. (In the absence of resetting, this approximation is valid because the fast system rapidly converges to a stable fixed point of the $x$ dynamics  
for fixed $y=\bary$.) Instead, we assume an averaging theorem holds so that we can replace $x(t)$ in the slow dynamics by its time average with respect to the fast stochastic process:
\begin{equation}
\frac{dy}{d\tau}=x_{\rm av}-y,\quad x_{\rm av}=\lim_{T\rightarrow \infty} \frac{1}{T}\int_0^Tx(t)dt.
\end{equation}
However, as it stands, it is not clear how to calculate $x_{\rm av}$. Therefore, we make a further assumption that the layer dynamics with resetting is ergodic in the sense that for fixed $y=\bary$ the time average is equivalent to an ensemble average with respect to the corresponding NESS (\ref{pF}): 
\begin{equation}
x_{\rm av}=\E[x|\bary]:=\int_{\R}xp^*(x|\bary) dx,
\end{equation}
Replacing $\bary$ by the slow variable $y(\tau)$ yields the averaged slow system
\begin{align}
\frac{dy}{d\tau}&=\E[x|y(\tau)]-y(\tau).
\label{sfav}
\end{align}
Finally, the solution of equation (\ref{sfav}) determines the slow variation of the NESS, that is, $p^*(x,\tau)=p^*(x|y(\tau))$.

\setcounter{equation}{0}
\section{FitzHugh-Nagumo model with resetting}

\begin{figure}[b!]
\centering
\includegraphics[width=12cm]{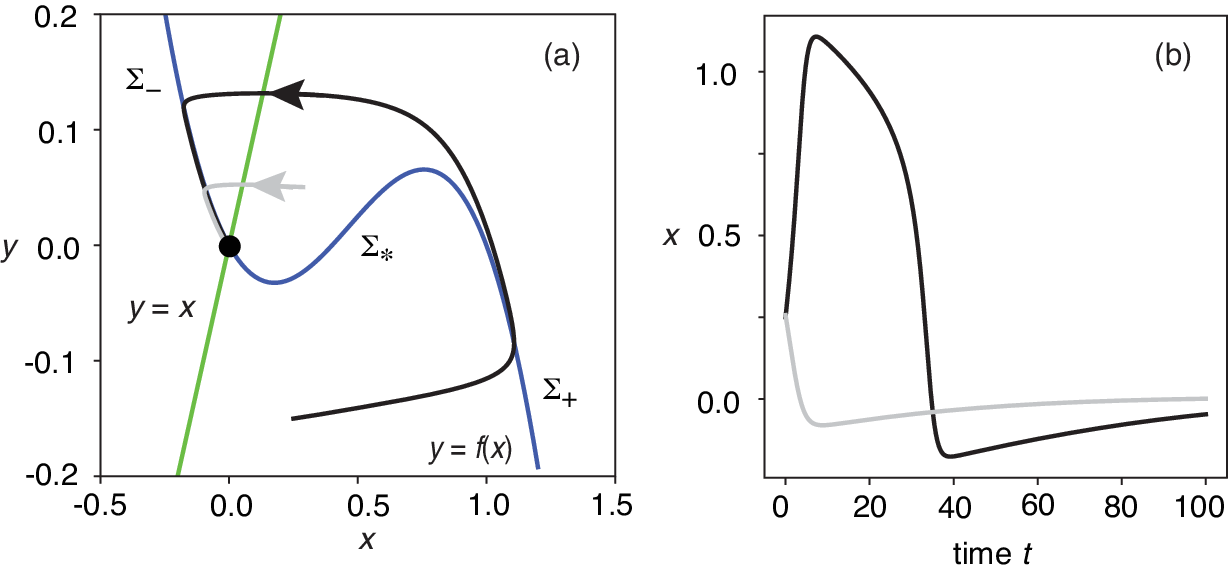} 
\caption{FN model in the excitable regime. (a) Nullclines $y=f(x)$ and $y=x$ for $a=0.4$ and $I_0=0$. In the excitable regime there exists a unique, stable fixed point on the left-hand branch. We also show a pair of trajectories with initial conditions on either side of the middle branch. The left, middle and right branches of the cubic nullcline are labeled by $\Sigma_-,\Sigma_*$ and $\Sigma_{+}$, respectively.(b) The corresponding plots of $x(t)$.} 
\label{fig3}
\end{figure}

\begin{figure}[b!]
\centering
\includegraphics[width=12cm]{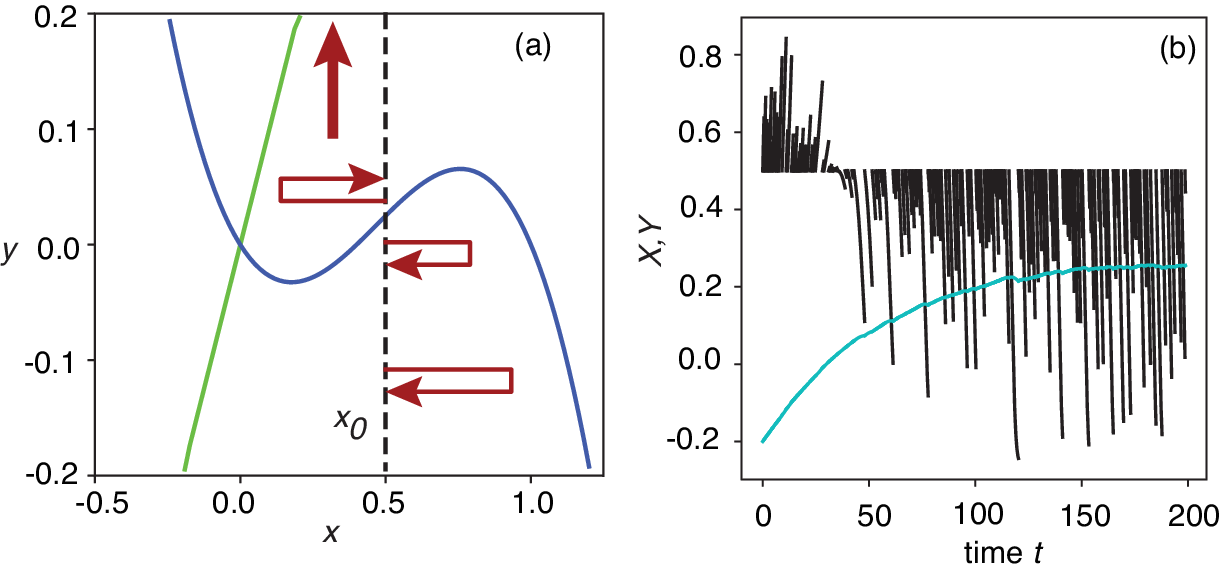} 
\caption{FN model with resetting in the excitable regime. (a) The fast dynamics with resetting is restricted to an interval that depends on the reset position $x_0$ and the current state of the slow variable $y$. Initially the fast dynamics has support on $[x_0,x_+)$ with $x_+\in \Sigma_+$. However, as the slow variable $y$ increases (indicated by thick vertical arrow), the system crosses the middle branch of the cubic nullcline and the support of the fast dynamics switches to $(x_-,x_0]$ with $x_-\in \Sigma_-$ (b) Numerical solution of equations (\ref{fsreset}) and (\ref{fC}) with $\epsilon=0.01$, $a=0.4$, $I_0=0$, $r=1$, $x_0=0.5$, $y_0=-0.2$ and. Black (dark) curves show $x(t)$ and blue (light) curves show $y(t)$. The solution exhibits the jump in the support of the fast variable and the saturation of the slow variable.} 
\label{fig4}
\end{figure}

\begin{figure}[b!]
\centering
\includegraphics[width=12cm]{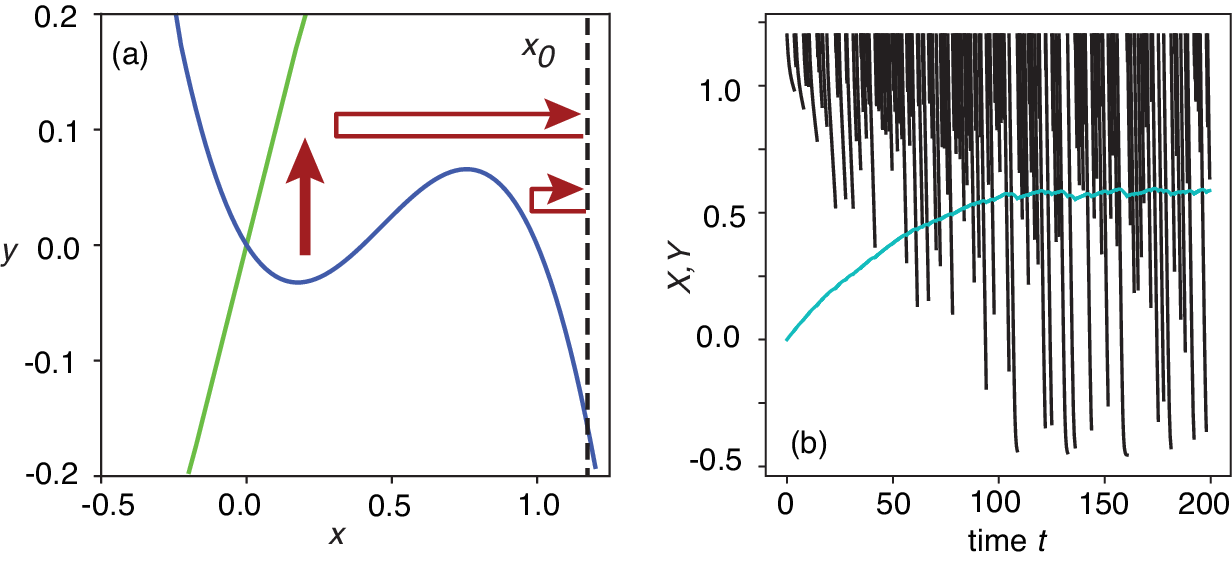} 
\caption{Same as Fig. \ref{fig4} except that $x_0=1.2,y_0=0$.} 
\label{fig5}
\end{figure}

\begin{figure}[t!]
\centering
\includegraphics[width=12cm]{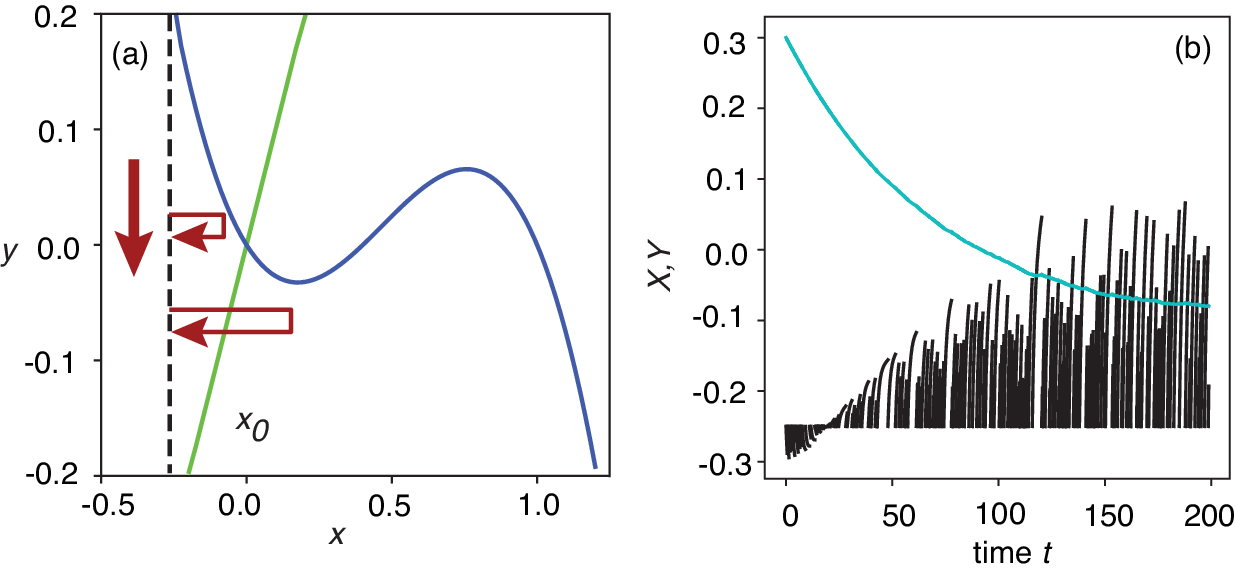} 
\caption{Same as Fig. \ref{fig4} except that $x_0=-0.25,y_0=0.3$.} 
\label{fig6}
\end{figure}

We  now use the separation of time scales outlined in \S 2 to analyze equations (\ref{fsreset})
  for $f(x)$ given by the cubic nonlinearity
\begin{align}
f(x)&= x(a-x)(x-1)+I_0,\quad 0<a<1,
\label{fC}
\end{align}
In the absence of resetting, equations (\ref{fsreset}) reduce to the classical FN model operating in an excitable or oscillatory regime \cite{Terman}. (Modifying the slow dynamics by taking $g(x,y)=x- by$ for $b \gg 1$ would also allow for a bistable regime.) 
We mainly focus on the excitable regime, but briefly discuss the oscillatory case at the end of the section. If $I_0=0$ in equation (\ref{fC}), then there exists a unique, stable fixed point on the left-hand branch of the cubic nullcline, see Fig. \ref{fig3}(a). Trajectories starting close to the fixed point remain in a neighborhood of the fixed point, whereas initial conditions located to the right of the middle branch make large excursions before returning to the fixed point. Within the context of conductance-based neural modeling, the large excursion is interpreted as an action potential, see Fig. \ref{fig3}(b).

As explained more generally in \S 2, the fast dynamics with resetting is restricted to an interval ${\mathcal I}(x_0,y)$ that depends on the reset position $x_0$ and the current value of the slow variable $y$. Suppose that the system starts in a state $(x_0,y_0)$ below the middle branch $\Sigma_*$, see Fig. \ref{fig4}. Initially the fast variable increases towards the right-hand branch $\Sigma_+$ after each reset while $y$ hardly changes.  Since the system is below the $y$-nullcline, $y$ slowly increases, resulting in a discontinuous jump of ${\mathcal I}(x_0,y)$ when $(x_0,y)$ crosses $\Sigma_*$. The fast variable then decreases towards the left-hand branch $\Sigma_-$ after each reset. The analogous scenarios when $x=x_0$ crosses the branches $\Sigma_{\pm}$ are shown in Figs. \ref{fig5} and \ref{fig6}, respectively. In all three cases we find that the slow variable converges to an asymptotic value $y^*$ that depends on $r$ and $x_0$. 
In the following we use a slow-fast analysis to understand this general result, which holds provided that $0<\epsilon \ll 1$ and $r=O(1)$.

\subsection{Fast dynamics with stochastic  resetting}

We begin by deriving an explicit expression for the NESS (\ref{pF}) using the factorization
\begin{equation}
f(x)-\bary=-(x-x_+(\bary))(x-x_*(\bary))(x-x_-(\bary)).
\label{roots}
\end{equation}
Suppose that the minimum and maximum of the cubic nullcline occur at the points $(x_{\min},y_{\min})$ and $(x_{\max},y_{\max})$, respectively.
If $y_{\min} < \bary < y_{\max}$ then the roots are real with $x_{\pm } \in \Sigma_{\pm}$, $x_*\in \Sigma_*$ and $x_-<x_*<x_+$. On the other hand, if $\bary>y_{\max}$ then $x_-\in \Sigma_-$ is real, whereas $x_*,x_+$ form a complex conjugate pair which we denote by $\mu \pm i\omega$.
Similarly, if $\bary < y_{\rm min}$ then $x_+\in \Sigma_+$ is real and $x_*,x_-$ are given by a complex conjugate pair $\mu'\pm i\omega'$. Substituting the factorized form into the integral of equation (\ref{Fir}) gives
\begin{align}
\int^x \frac{ds}{f(s)-\bary} &=
\int^x  \frac{ ds}{(x_+-s)(s-x_*)(s-x_-)}
\end{align}
In order to evaluate this integral we use partial fractions. However, the details will depend on how many roots are real.
\medskip

\noindent {\bf Case I: $y_{\min} < \bary<y_{\max}$.} Writing
\begin{equation}
 \frac{ 1}{(s-x_+)(s-x_*)(s-x_-)}=\frac{C_+}{x_+-s}+\frac{C_*}{s-x_*}+\frac{C_-}{s-x_-}
 \end{equation}
 and comparing both sides implies that
 \begin{equation}
 C_+(s-x_*)(s-x_-)+ C_*(s-x_+)(s-x_-)+ C_-(s-x_*)(s-x_+)=1.
 \end{equation}
 Collecting terms in powers of $s$ yields the simultaneous equations
 \begin{subequations}
 \label{C}
 \begin{align}
 &C_++C_*+C_-=0\\
 &C_+(x_*+x_-)+C_*(x_++x_-)+C_-(x_*+x_+)=0\\
 &C_+x_*x_- + C_*x_+x_- +C_-x_+x_*=1.
 \end{align}
\end{subequations}
Setting $C_*=-C_+-C_-$ in equation (\ref{C}b) implies that
$C_-=C_+(x_+-x_*)/(x_*-x_-)$.
Finally, substituting for $C_-,C_*$ in equation (\ref{C}c) determines $C_+$ such that
\begin{align}
C_+&=\frac{1}{(x_+-x_-)(x_+-x_*)},\quad C_-=\frac{1}{(x_+-x_-)(x_*-x_-)},\nonumber \\
C_*&=-\frac{1}{(x_+-x_*)(x_*-x_-)}.
\label{Cs}
\end{align}

\begin{figure}[t!]
\centering
\includegraphics[width=12cm]{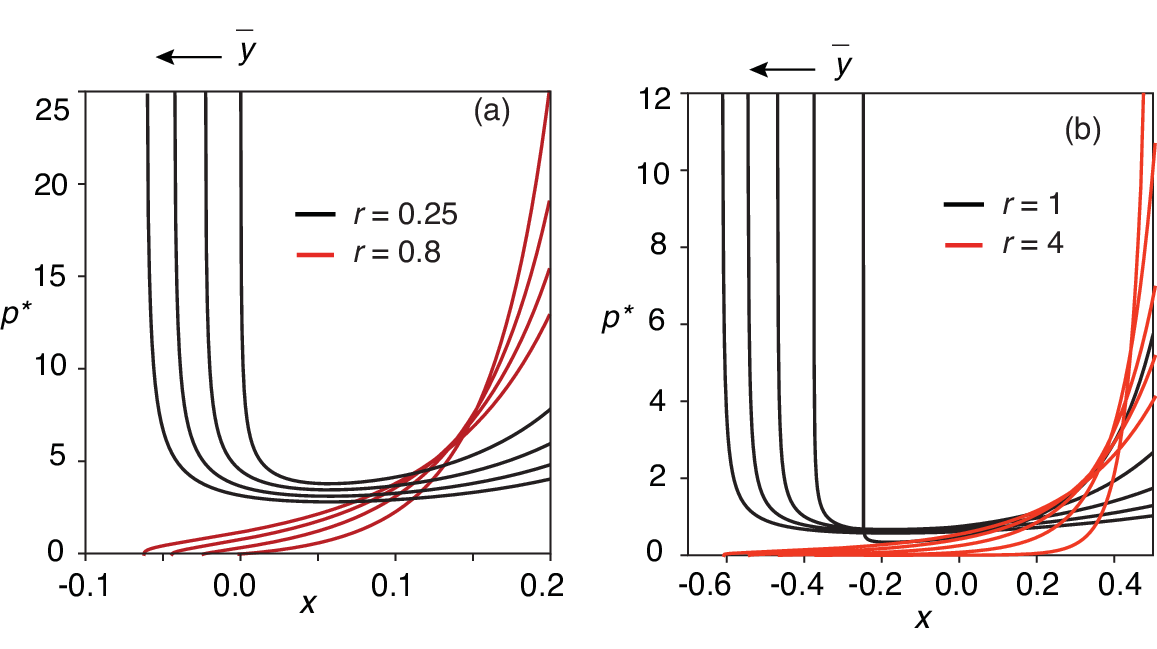} 
\caption{Plots of $p^*(x)$ as a function of the fast variable $x$ for the FN model with resetting in the excitable regime ($I_0=0,a=0.4$). (a) Case I: $f(x)-\bary$ has 3 real roots $x_{\pm},(\bary) ,x_*(\bary) $ with $\bary=0.0,0.01,0.02.0.03$, and $p(x)$ given by equation (\ref{p1}). We take $x_0=0.2$ and $r=0.25$ (black curves) or $r=0.8$ (red curves).  (b)  Case II: one real root $x_-(\bary)$ with $\bary=0.2,0.4,0.6.0.8,1.0 $ and $p^*(x)$ given by equation (\ref{ppp}). We take $x_0=0.5$, and $r=1$ (black curves) or $r=4$ (red curves).} 
\label{fig7}
\end{figure}

It now follows that
 \begin{align}
\int^x \frac{ds}{f(s)-\bary} =
-C_+\ln |x -x_+|-C_-\ln |x -x_-|-C_*\ln |x -x_*|.
\end{align}
We thus obtain the result
\begin{equation}
\Gamma(x)=|x-x_+|^{\beta_+}|x-x_-|^{\beta_- }|x-x_*|^{-\beta_*}  
\label{gamlog}
\end{equation}
with 
\begin{equation}
\beta_k =r|C_k|,\quad k=\pm, *.
\label{bk}
\end{equation}
The support of the NESS will depend on the location of $x_0$. For example, if $x_-< x_0<x_*$ then the state is restricted to the subinterval  $x(t)\in (x_-,x_0]$ and $A_+=0$ in equation (\ref{pF}). This implies that 
\begin{align}
p^*(x)&=\frac{r }{(x_+-x)(x_*-x)(x-x_-)}\nonumber  \\
&\quad \times  \left  ( \frac{x_+-x}{x_+-x_0}\right )^{\beta_+}\left (\frac{x-x_-}{x_0-x_-}\right )^{\beta_-} \left (\frac{x_*-x}{x_*-x_0}\right )^{-\beta_*},
\label{p1}
\end{align}
which is singular as $x\rightarrow x_-$ from above when $\beta_- <1$. On the other hand, if $x_*< x_0<x_+$ then the state is restricted to the subinterval  $x(t)\in [x_0,x_+)$ and $A_-=0$ in equation (\ref{pF}). The NESS now takes the form
\begin{align}
p^*(x)&=\frac{r }{(x_+-x)(x-x_*)(x-x_-)}\nonumber  \\
&\quad \times  \left  ( \frac{x_+-x}{x_+-x_0}\right )^{\beta_+}\left (\frac{x-x_-}{x_0-x_-}\right )^{\beta_-} \left (\frac{x-x_*}{x_0-x_*}\right )^{-\beta_*},
\label{p2}
\end{align}
which is singular as $x\rightarrow x_+$ from below when $\beta_+<1$. The cases $x_0<x_-$ and $x_0>x_+$ can be handled in a similar fashion. 
Example plots of $p^*(x)$ given by equation (\ref{p1}) are shown in Fig. \ref{fig7}(a) for $\bary \in [0,y_{\max}]$. It can be checked numerically that $\int_{x_-}^{x_0} p^*(x)dx=1$ in the given parameter regime.
\medskip

\noindent{\bf Case II: $ \bary > y_{\max}$.} Setting $x_+=\mu+i\omega$ and $x_*=\mu-i\omega$, we have
\begin{equation}
 (s-x_+)(s-x_*)(s-x_-)=(s-x_-)([s-\mu]^2+\omega^2).
  \end{equation}
  The explicit values of $x_-$, $\mu$ and $\omega$ as a function of $\bary$ are obtained from the general cubic formula in appendix B.
  The partial fraction decomposition takes the form
\begin{align}
\frac{1}{(s-x_+)(s-x_*)(s-x_-)}=\frac{C_-}{s-x_-}+\frac{C(s-\mu)+D} {[s-\mu]^2+\omega^2}.
\label{fact}
\end{align}
Comparing both sides implies that
 \begin{equation}
C_-(s^2-2\mu s+\mu^2+\omega^2)+C(s^2-(\mu+x_-)s+\mu x_-)+D(s-x_-)=1.
 \end{equation}
 Collecting terms in powers of $s$ gives $C=-C_-$ and
 \begin{subequations}
 \label{CD}
 \begin{align}
 &C_-(x_--\mu)+D =0\\
 &C_-(\mu^2+\omega^2-\mu x_-)-Dx_-=1.
 \end{align}
\end{subequations}
We thus find that
\begin{align}
C_-=\frac{1}{(\mu-x_-)^2+\omega^2},\quad C=-C_-,\quad  D=(\mu-x_-)C_-.
\end{align}
Substituting equation (\ref{fact}) into the integral of equation (\ref{Fir}) gives
\begin{align}
\int^x \frac{ds}{f(s)-\bary} =
-C_-\ln |x -x_-| - \frac{C}{2} \ln [(x-\mu)^2+\omega^2] -\frac{D}{\omega}\tan^{-1}\left (\frac{x-\mu}{\omega}\right )
\end{align}
We thus obtain the general result
\begin{equation}
\Gamma(x)=|x-x_-|^{\beta_-}  [(x-\mu)^2+\omega^2]^{-\beta_-/2}\exp\left (\frac{rD}{\omega}\tan^{-1}\left (\frac{x-\mu}{\omega}\right )\right ).
\end{equation}
with  $\beta_-=rC_-$. Since $x(t)\in [x_-,x_0]$, $A_+=0$ and the NESS is
\begin{align}
p^*(x)&=\frac{r }{(x-x_-)[(x-\mu)^2+\omega^2]} \nonumber  \\
&\quad \times  \left (\frac{x-x_-}{x_0-x_-}\right )^{\beta_-} \left (\frac{(x-\mu)^2+\omega^2}{(x_0-\mu)^2+\omega^2}\right )^{-\beta_-/2}\nonumber \\
&\quad \times  \exp\left (\frac{rD}{\omega}\tan^{-1}\left (\frac{x-\mu}{\omega}\right )\right )\exp\left (-\frac{rD}{\omega}\tan^{-1}\left (\frac{x_0-\mu}{\omega}\right )\right ).
\label{ppp}
\end{align}
Example plots of $p^*(x)$ are shown in Fig. \ref{fig7}(b). Again it can be checked that $\int_{x_-}^{x_0} p^*(x)dx=1$.

\subsection{Slow dynamics and averaging}
So far we have focussed on the fast dynamics with resetting for fixed $y=\bary$. In particular, we calculated the NESS $p^*(x|\bary)$ which depended on $\bary$ via the roots $x_{\pm}(\bary),x_*(\bary)$ of equation (\ref{roots}). Replacing $\bary$ by the slow variable $y(\tau)$, we obtain the averaged equation (\ref{sfav}). In Figs. \ref{fig4}--\ref{fig6} we found that $y(\tau)\rightarrow y^*$ in the limit $\tau\rightarrow \infty$. This general result follows from the following observations:
\medskip

\noindent i)  The stable branches $\Sigma_{\pm}$ of the cubic nullcline have negative slope, that is,
\begin{equation*}
\frac{dx_{\pm}(y)}{dy}<0\mbox{ for all } y \in \R.
\end{equation*}

\noindent ii) The support ${\mathcal I}(x_0,y)$ of $p^*(x|y)$ belongs to the  interval set
\[\{(x_-(y), x_0],\ [x_0,x_-(y)),\ [x_0,x_+(y)),\ (x_+(y),x_0])\}.
\]
In all cases the interval shifts to more negative values of $x$ as $y$ increases for fixed $x_0$.
\smallskip

\noindent iii) Hence, the expectation $x_{\rm av}(y)$ is a monotonically decreasing function of $y$, since
\[x_{\rm av}(y)=\int_{{\mathcal I}(x_0,y)}xp^*(x|y)dx.\]

\noindent iv) If $y(0)< x_{\rm av}(y(0))$ then $y(\tau)$ increases according to equation (\ref{sfav}) which reduces $x_{\rm av}$. This occurs in Figs. \ref{fig4} and \ref{fig5}. Similarly, if $y(0)> x_{\rm av}(y(0))$ then $y(\tau)$ decreases and $x_{\rm av}$ increases, see Fig. \ref{fig6}.
\medskip

\noindent Therefore, taking the large time limit of equation (\ref{sfav}) yields a self-consistency condition for $y^*$:
\begin{equation}
y^*=\E[x|y^*]=\int_{{\mathcal I}(x_0,y^*)}xp^*(x|y^*)dx.
\end{equation}
Note that the asymptote depends on the resetting rate $r$ and the resetting point $x_0$.
 
\begin{figure}[t!]
\centering
\includegraphics[width=12cm]{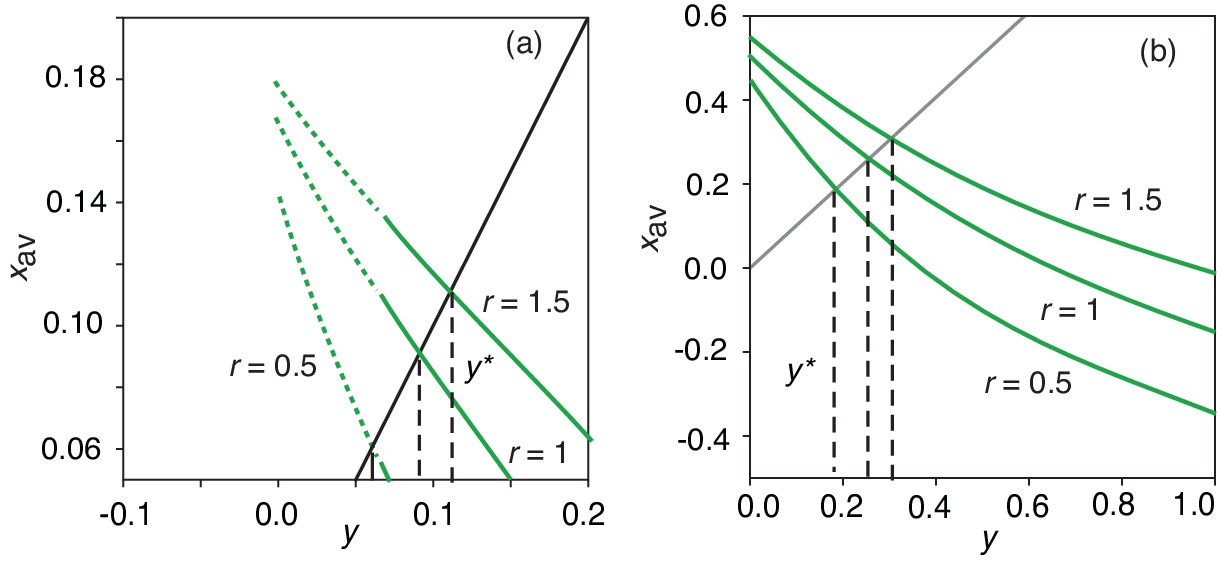} 
\caption{FN model with resetting in the excitable regime. Numerical plots of $x_{\rm av}=\E[x|y]$ as a function of $y$ for $r=0.5,1.0,1.5$. In each case the intercept with the straight line $x_{\rm av}=y$  determines the stable fixed point $y^*$ of the slow variable. Dashed green curves indicate the existence of 3 real roots and $p^*(x)$ given by equation (\ref{p1}), whereas solid curves indicate the existence of 1 real toot and $p^*(x)$ given by equation (\ref{ppp}).  (a) $x_0=0.2$. (b) $x_0=0.5$.} 
\label{fig8}
\end{figure}

In Fig. \ref{fig8} we plot $x_{\rm av}=\E[x|y] $ as a function of $y$ for various values of the resetting rate $r$ and reset position $x_0$. 
In each case there is a unique intercept with the straight line $x_{\rm av}=y$, which we identify with the asymptote $y^*$. Moreover, it can be checked that the averaged equation is in good agreement with the full dynamics on long time scales. This can be seen explicitly for the specific example $r=1$ and $x_0=0.5$ by comparing $y^*$ in Fig. \ref{fig8}(b) with the asymptote in Fig. \ref{fig4}(b).
Finally, note that the NESS of the fast variable also reaches a steady-state in the long-time limit,
\begin{equation}
\lim_{\tau\rightarrow \infty}p^*(x|y(\tau))=p^*(x|y^*).
\end{equation}

\begin{figure}[t!]
\centering
\includegraphics[width=11cm]{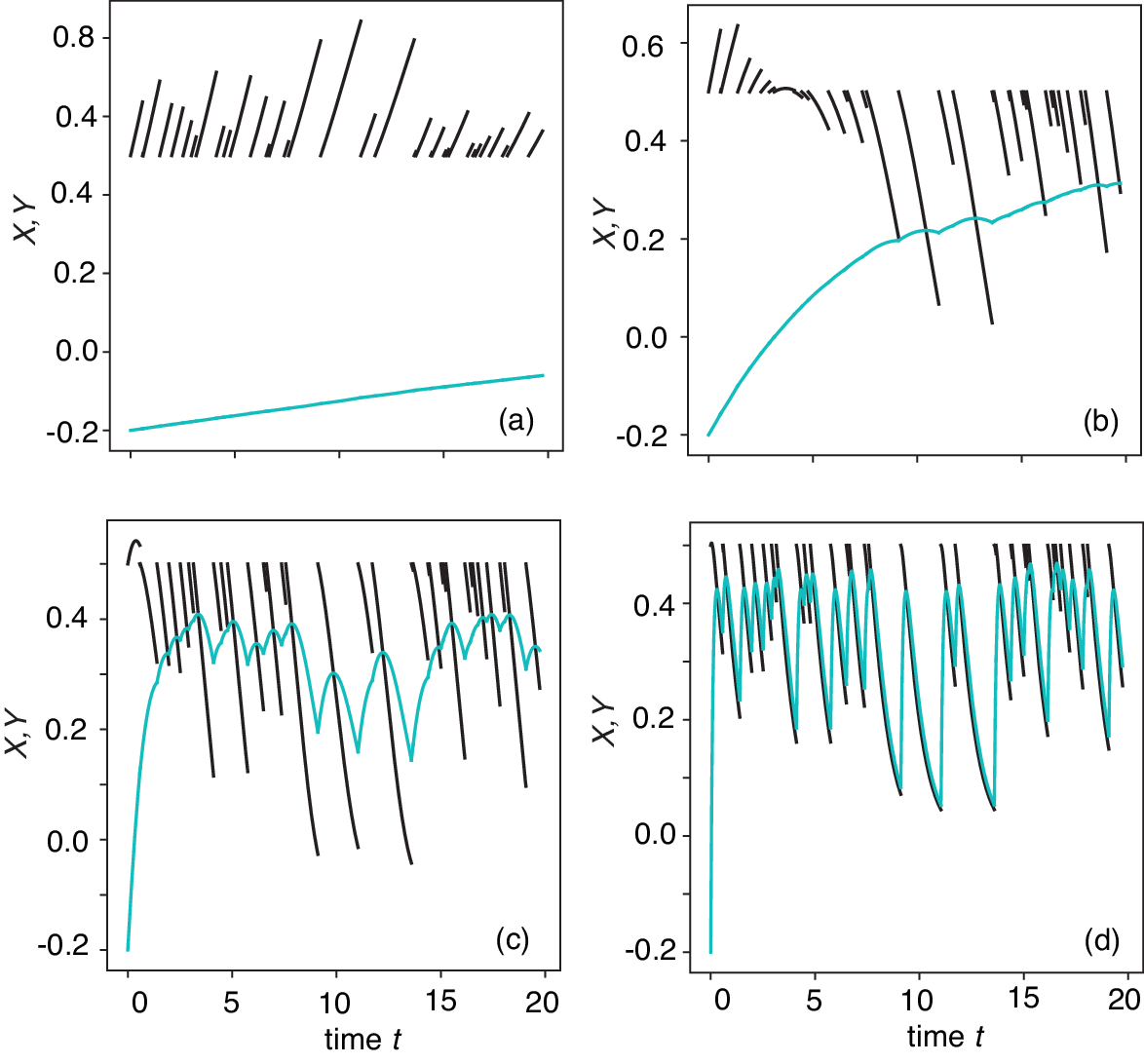} 
\caption{FN model with resetting in the excitable regime. Numerical solutions of equations (\ref{fsreset}) and (\ref{fC}) for different values of $\epsilon$: (a) $\epsilon=0.01$, (b) $\epsilon=0.1$, (c) $\epsilon =1$, (d) $\epsilon =10$. Other parameters are $r=1$, $x_0=0.5$, $y_0=-0.2$ and $a=0.4$. Black (dark) curves show $x(t)$ and blue (light) curves show $y(t)$.} 
\label{fig9}
\end{figure}

\begin{figure}[t!]
\centering
\includegraphics[width=11cm]{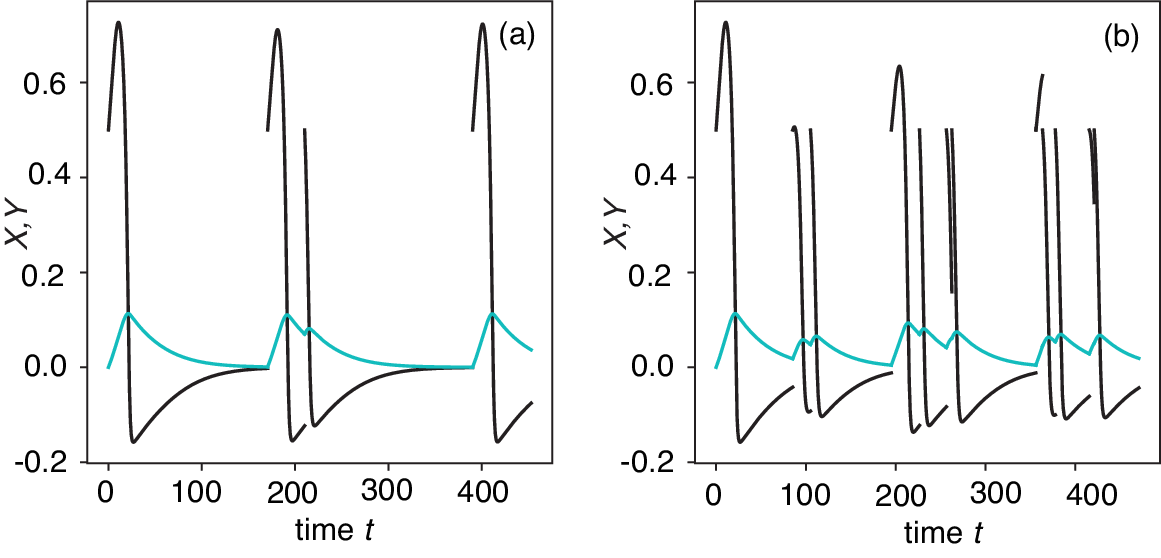} 
\caption{Numerical solutions of equation (\ref{fsreset}) with resetting for $f(x)=x(x-a)(1-x)$ and slow resetting: (a) $r=0.02$ and (b) $r=0.04$. Other parameters are $\epsilon=0.01$, $x_0=0.5$, $y_0=0.0$ and $a=0.4$. Black (dark) curves show $x(t)$ and blue (light) curves show $y(t)$.} 
\label{fig10}
\end{figure}

There are two crucial assumptions regarding the separation of time scales used in the above analysis. First, $\epsilon \ll 1$ and second $r=O(1)$. The latter ensures that resetting occurs on time-scales consistent with the fast system. In Fig. \ref{fig9} we illustrate how the separation of time scales for small $\epsilon$ breaks down as $\epsilon$ increases for fixed $r$. Note, in particular, that for sufficiently large $\epsilon$, $y(t)$ tracks $x(t)$. On the other hand, if $\epsilon \ll 1$ and $r=O(\epsilon)$ then there is still a fast-slow separation but the dynamics of the slow variable is no longer determined by the distribution of the fast variable with respect to an NESS. This is illustrated in Fig. \ref{fig10} for $x_0=0.5$, which shows that almost every reset event triggers an ``action potential'' .

 \subsection{Oscillatory regime ($I_0=0.4$)}

\begin{figure}[t!]
\centering
\includegraphics[width=12cm]{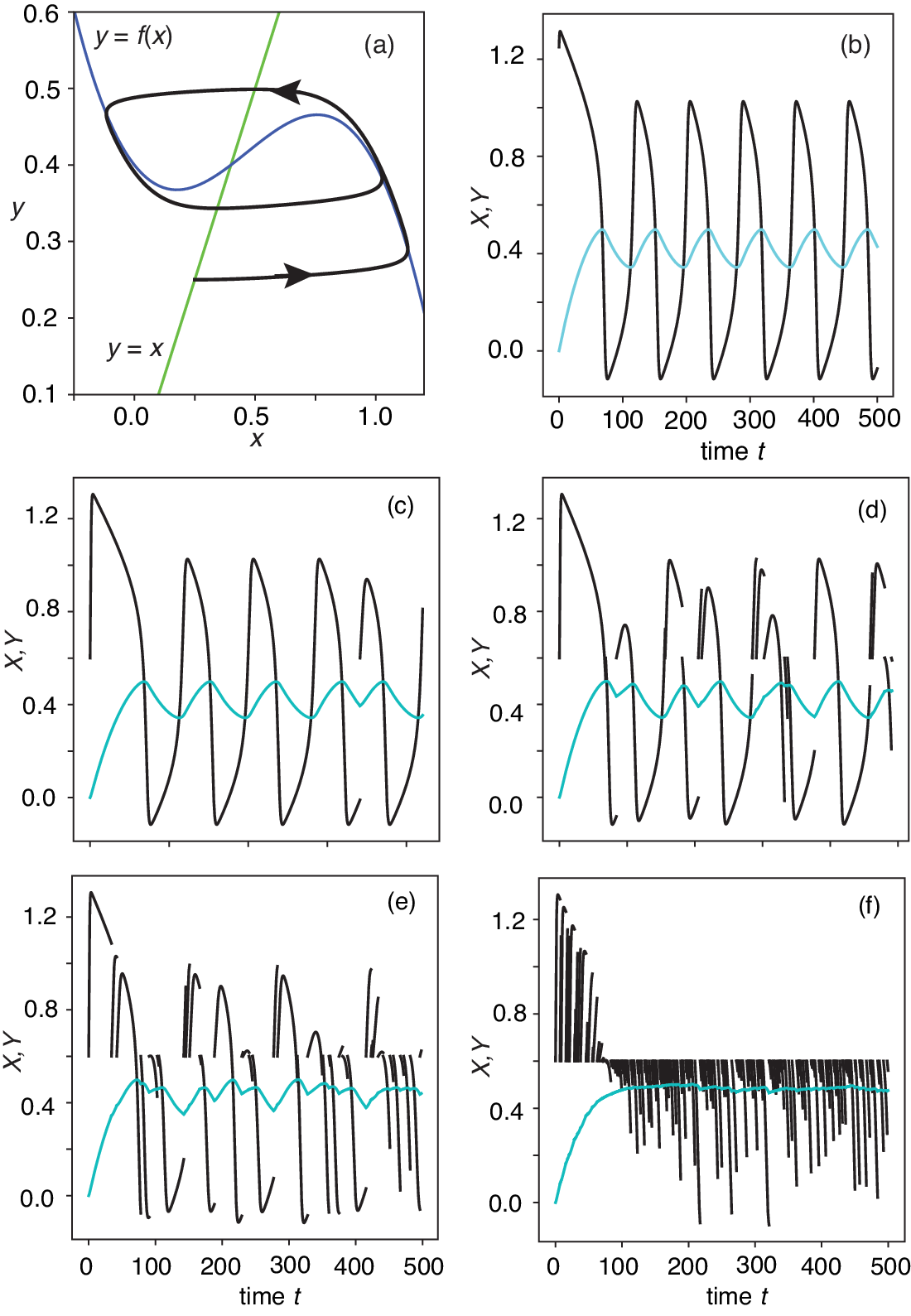} 
\caption{FN model in the oscillatory regime. (a) Nullclines $y=f(x)$ and $y=x$ for $f(x)=x(x-a)(1-x)+I_0$, $a=0.4$ and $I_0=0.4$. In the absence of resetting there exists a unique, unstable fixed point on the middle branch surrounded by a stable limit cycle. Also shown is a sample trajectory for $\epsilon=0.01$. (b) Plots of $x(t),y(t)$ as a function of time $t$ for the trajectory shown in (a). (c-f) Numerical solutions of the FN equation with resetting for (c) $r=0.01$, (d) $r=0.05$, (e) $r=0.1$ and (f) $r=0.5$, respectively.}
\label{fig11}
\end{figure}

As our final example, suppose that  in the absence of resetting, equations (\ref{fsreset}) and (\ref{fC}) reduce to the FN model operating in the oscillatory regime. The fixed point now lies on the middle branch $\Sigma_*$ of the cubic nullcline and is unstable. In the oscillatory regime there exists a stable limit cycle surrounding the unstable fixed point as illustrated in Fig. \ref{fig11}(a,b). Within the context of conductance-based neural modeling, the output of the model is a periodic sequence of action potentials. As might be expected, if resetting occurs at a rate consistent with the fast dynamics, then oscillations are eliminated and we can apply the slow-fast analysis of previous sections. On the other hand, if $r=O(\epsilon)$ then resetting results in noisy oscillations. The change in behavior as a function of $r$ is illustrated in Fig. \ref{fig11}(c-f).

\setcounter{equation}{0}
\section{Accumulation time of the fast process} 

Our analysis of the fast dynamics with resetting assumed that the probability density $p(x,t)$ converged sufficiently quickly to the NESS $p^*(x)$ so  that we could average the slow dynamics with respect to $p^*(x)$. The characteristic time-scale for the relaxation to steady-state is $r^{-1}$. Numerically, we find that averaging theory holds provided that $r\gg \epsilon$. In this section we explore another measure of the approach to steady state based on the notion of an accumulation time. The latter was originally developed within the context of diffusion-based morphogenesis \cite{Berez10,Berez11,Berez11a}, but has more recently been applied to intracellular protein gradient formation \cite{Bressloff19} and to diffusion processes with stochastic resetting \cite{Bressloff21}. 

Following Ref. \cite{Bressloff21}, we first decompose the solution of equation  (\ref{Liou}) as $p(x,t)=p_a(x,t)+p_d(x,t)$ with
\begin{subequations}
\label{pad}
\begin{align}
\frac{\partial p_a(x,t)}{\partial t}&=-\frac{\partial [(f(x)-\bary)p_a(x,t)]}{\partial x}-rp_a(x,t)+r\delta(x-x_0),\quad  p_a(x,0)=0,\\
 \frac{\partial p_d(x,t)}{\partial t}&=-\frac{\partial [(f(x)-\bary)p_d(x,t)]}{\partial x}-rp_d(x,t),\quad  p_d(x,0)=\delta(x-x_0).
\end{align}
\end{subequations}
That is, $p_a(x,t)$ represents the accumulating component of the probability density that grows from zero and approaches the NESS $p^*(x)$ as $t\rightarrow \infty$, whereas $p_d(x,t)$ is the complementary decreasing component that starts with its mass at $x=0^+$ and decays to zero as $t\rightarrow \infty$. For concreteness, suppose that the support of $p(x,t)$ is $[x_0,x_+)$ where $x_+$ is a stable fixed point of the layered problem without resetting. Integrating equations (\ref{pad}) with respect to $x\in [x_0,x_+)$ implies that
\begin{equation}
\frac{dQ_a}{dt}=r-rQ_a(t),\quad Q_a(0)=0; \quad \frac{dQ_d}{dt}= -rQ_d(t), \quad Q_d(0)=1,
\end{equation}
with
\begin{equation}
Q_a(t)=\int_{x_0}^{x_+}p_a(x,t)dx,\quad Q_a(t)=\int_{x_0}^{x_+}p_a(x,t)dx.
\end{equation}
It follows that
\begin{equation}
Q_a(t)=1-\e^{-rt},\quad Q_d(t)=\e^{-rt}.
\label{QA}
\end{equation}
with $Q_a(t)+Q_d(t)=1$ as expected.

Introduce the function
\begin{equation}
\label{accu}
Z(x,t)=1-\frac{p_a(x,t)}{p^*(x)},
\end{equation}
which represents the fractional deviation of the probability density $p_a(x,t)$ from the steady-state. Assuming there is no overshooting then, $1-Z(x,t)$ can be interpreted as the fraction of the steady-state density that has accumulated at $x$ by time $t$. It follows that $-\partial_t Z(x,t)dt$ is the fraction accumulated in the interval $[t,t+dt]$. The accumulation time is then defined by analogy to mean first passage times \cite{Berez10,Berez11,Berez11a}:
\begin{equation}
\label{accu2}
T(x)=\int_0^{\infty} t\left (-\frac{\partial Z(x,t)}{\partial t}\right )dt=\int_0^{\infty} Z(x,t)dt.
\end{equation}
It is usually more useful to calculate the accumulation time in Laplace space. 
Laplace transforming equation (\ref{accu}) and setting ${\widetilde{F}(x,s)}=s{\widetilde{p}_a(x,s)}$ gives
\[s\widetilde{Z}(x,s)=1-\frac{\widetilde{F}(x,s)}{p^*(x)},\quad \lim_{s\rightarrow 0}\widetilde{F}(x,s)=p^*(x)\]
and, hence
\begin{eqnarray}
 T(x)=\lim_{s\rightarrow 0} \widetilde{Z}(x,s) = \lim_{s\rightarrow 0}\frac{1}{s}\left [1-\frac{\widetilde{F}(x,s)}{p^*(x)}\right ] =-\frac{1}{p^*(x)}
\left .\frac{d}{ds}\widetilde{F}(x,s)\right |_{s=0}.
\label{acc}
\end{eqnarray}

The next step is to Laplace transform equation (\ref{pad}a) with $p_a(x,0)=0$:
\begin{equation}
\label{pLT}
 s\widetilde{p}_a(x,s) =-\frac{\partial [(f(x)-\bary)\widetilde{p}_a(x,s)]}{\partial x}-r \widetilde{p}_a(x,s)+\frac{r}{s}\delta(x-x_0).
\end{equation}
Multiplying both sides by $s$ and taking the limit $s\rightarrow 0$ with
\[p^*(x)=\lim_{t\rightarrow \infty} p_a(x,t)=\lim_{s\rightarrow 0}s\widetilde{p}_a(x,s),\]
recovers equation (\ref{NESS}) for $p^*(x)$.
On the other hand, multiplying both sides of equation (\ref{pad}a) by $-s$, differentiating with respect to $s$ and then taking the limit $s\rightarrow 0$ gives
\begin{eqnarray}
\frac{\partial [(f(x)-\bary)T(x)p^*(x)]}{\partial x}+r p^*(x)T(x)=p^*(x) 
\end{eqnarray}
Setting $\tau(x)=[f(x)-\bary]T(x)p^*(x)$ gives
\begin{eqnarray}
\frac{\partial \tau(x)]}{\partial x}+\frac{r\tau(x)}{f(x)-\bary}=p^*(x), 
\end{eqnarray}
which can be solved using the integrating factor $\Gamma(x)^{-1}$ with $\Gamma(x)$ defined by equation (\ref{Fir}). That is,
\begin{align}
\frac{d}{dx}\left [\frac{\tau(x)}{\Gamma(x)}\right ] =\frac{p^*(x)}{\Gamma(x)}.
\end{align}
For concreteness, suppose that the support of $p^*(x)$ is $[x_0,x_+)$ so that equation (\ref{pstar0}) holds. Integrating with respect to $x$ then yields
\begin{equation}
\frac{\tau(x)}{\Gamma(x)}-\frac{\tau(x_0)}{\Gamma(x_0)}=-\frac{1}{\Gamma(x_0)}\int_{x_0}^x \frac{\Gamma'(z)}{\Gamma(z)}dz=\frac{\log \Gamma(x_0)-\log \Gamma(x)}{\Gamma(x_0)}.
\end{equation}
Finally, noting that
\begin{equation}
\frac{\tau(x)}{\Gamma(x)}=\frac{[f(x)-\bary]T(x)p^*(x)}{\Gamma(x)}=rT(x),
\end{equation}
we obtain the result
\begin{equation}
T(x)=T(x_0)+\frac{\log \Gamma(x_0)-\log \Gamma(x)}{r\Gamma(x_0)}.
\label{Tx}
\end{equation}

\begin{figure}[t!]
\centering
\includegraphics[width=12cm]{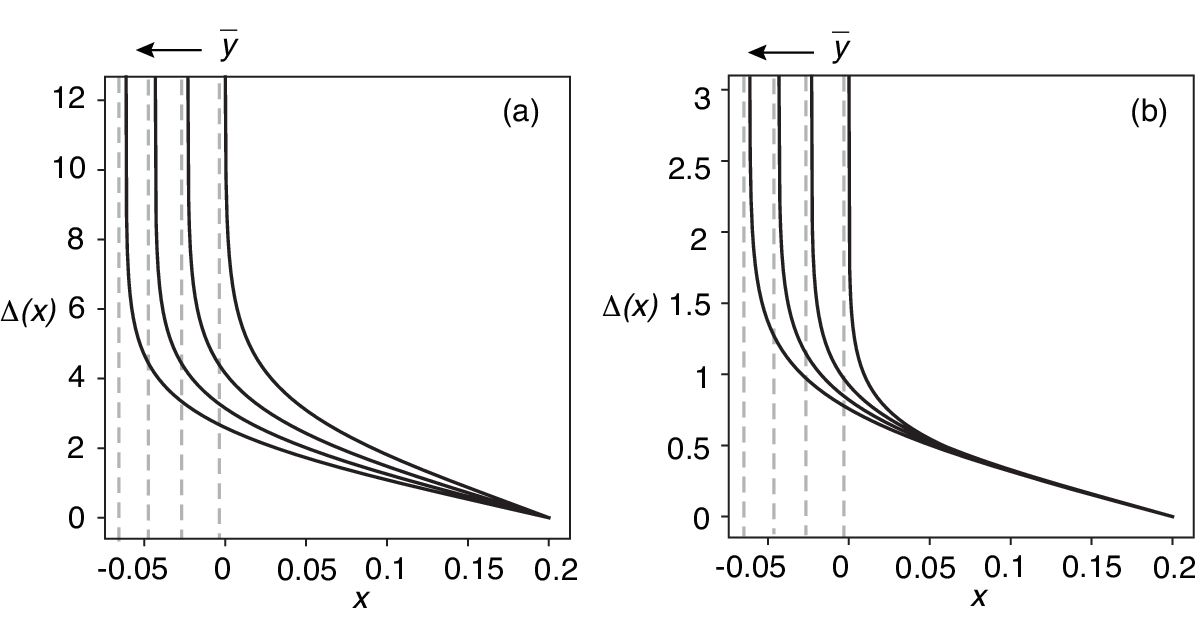} 
\caption{FN model with resetting in the excitable regime ($I_0=0,a=0.4$). Plots of the accumulation time $\Delta T(x)=T(x)-T_0(x)$ with $T(x)$ given by equation (\ref{Tx}). We consider the values $\bary=0,0.04,0.08,0.12$ (indicated by vertical dashed lines) with (a) $r=0.25$ and (b) $r=1$. We also take $x_0=0.2$.} 
\label{fig12}
\end{figure}

As a specific example, consider the FN model with $x_-<x_0<x_*$
 and $\Gamma(x)$ given by equation (\ref{gamlog}). Substituting for $\Gamma$ in equation (\ref{Tx}) gives
 \begin{align}
T(x)&= T(x_0)   +\bigg \{C_+ \log\left (\frac{x_+-x_0}{x_+-x}\right )+C_- \log\left (\frac{x_0-x_-}{x-x_-}\right )+C_*\log\left (\frac{x_*-x_0}{x_*-x}\right )\bigg \}\nonumber \\
&\qquad \qquad \qquad \times \frac{1}{(x_+-x_0)^{\beta_+}(x_0-x_-)^{\beta_- }(x_*-x_0)^{-\beta_*} }.
\end{align}
with $C_k$ and $\beta_k$ for $k=\pm,*$ defined in equations (\ref{Cs}) and (\ref{bk}), respectively.
Example plots of $\Delta T(x)\equiv T(x)-T(x_0)$ are shown in Fig. \ref{fig12}. We see that 
$\Delta T(x)$ is a positive monotonically increasing function of $x$, $x\in (x_-,x_0]$. The accumulation time effectively captures the additional time need to approach steady state as one moves away from the reset point $x_0$. Note that  there exists a boundary layer around $x=x_-(\bary)$ where the accumulation time blows up. This reflects the fact that between resets the system converges towards (but never reaches) the stable fixed point $x_-$. In summary, equations (\ref{QA}) and (\ref{Tx}) provide complementary representations of the relaxation process for the layer problem with resetting.

\setcounter{equation}{0}
\section{Discussion}  In this paper we explored the effects of instantaneous stochastic resetting on the dynamics of a planar slow-fast system in which only the fast variable resets. The inverse resetting rate $r^{-1}$ introduced an additional time-scale that had a major effect on the dynamics. If $r\gg \epsilon$ (fast resetting), then the slow dynamics reduced to an averaged equation that depended on the mean of the fast variable with respect to a corresponding NESS. Moreover, the solution of the averaged equation converged to an $r$-dependent fixed point. On the other hand, if $r=O(\epsilon)$ (slow resetting), then each resetting event triggered a trajectory that converged to an attractor of the underlying deterministic system. In the case of the FN model, the attractor was either a stable fixed point (excitable regime) or a stable limit cycle (oscillatory regime). Resetting triggered a sequence of pulses or action potentials in the excitable regime and fluctuations of the limit cycle in the oscillatory regime.

In contrast to most studies of stochastic resetting \cite{Evans20}, we considered the system without resetting to be a deterministic dynamical system rather than a stochastic differential equation. This was partly motivated by recent studies of the dynamics on the OA manifold of Kuramoto-type models with global stochastic resetting \cite{Sarkar22,Bressloff24a,Bressloff25}. A mathematical reason for ignoring the effects of diffusion is that the resulting parameterized NESS $p^*(x|y)$ maintains certain information about the underlying structure of the nullclines, in the sense that the support of $p^*(x|y)$ is from $x_0$ to a point on the nullcline given by a root of $y=f(x)$. Diffusion would result in a support on $\R$.

In future work it would be interesting to explore the effects of stochastic resetting on a wider class of slow-fast systems \cite{Wech20}. This could include higher-dimensional systems as well as non-standard forms that exhibit multiple time-scale dynamics, but can only be reduced
to the standard separable form locally. A classical example of the latter is a two-stroke relaxation oscillator \cite{Jelbart20}. However, restricting stochastic resetting to the fast component of the dynamics is more complicated. Finally, in our separation of time scales we assumed that averaging theory could be applied to the reduced slow system and checked its validity against numerical simulations. It would be useful to develop a more rigorous analysis by extending the machinery of geometric singular perturbation theory to include the effects of resetting.

\setcounter{equation}{0}
\renewcommand{\theequation}{A.\arabic{equation}}
\section*{Appendix A: Derivation of the modified Liouville equation}

In this appendix we derive the generalized Liouville equation (\ref{Liou}) from equation (\ref{xreset}) using the stochastic calculus of jump processes. For the sake of generality, we consider the stochastic differential equation (SDE)
\begin{equation}
dX(t)=F(x)dt+\sqrt{2D}dW(t)+[x_0-X(t^-)]dN(t),
\end{equation}
where $W(t)$ is a Brownian motion and $D$ is a diffusivity. Setting $D=0$ and $F(x)=f(x,\bary)$ recovers the differential version of equation (\ref{xreset}).  Let $u(x)$ be an arbitrary smooth bounded test function on $\R$. Using a generalized It\^o's lemma, the infinitesimal $du(X(t))$ can be decomposed as
\begin{eqnarray}
du(X(t))&=&[u'(X(t))F(X(t))+Dg''(X(t)]dt+\sqrt{2D}u'(X(t))dW(t)\nonumber \\
&&\quad +\bigg [u(x_0)-u(X(t^-))\bigg ]dN(t).
\label{lemma2}
\end{eqnarray}
Given an arbitrary test function $u$, we have the identity
\begin{eqnarray}
\left [  \int_{\R} u(x)\frac{\partial \rho(x,t)}{\partial t}  dx\right ]dt =du(X(t) ),
\label{ffr}
 \end{eqnarray}
 where $\rho(x,t)=\delta(x-X(t))$.
Applying It\^o's lemma (\ref{lemma2}) and equation (\ref{hN}) gives 
    \begin{eqnarray}
    \label{jar}
   \int_{\R}   u(x)\frac{\partial \rho(x,t)}{\partial t}  dx  
& = &\int_{\R} \rho(x,t) \bigg [  u'(x)F(x)+Dg''(x)+\sqrt{2D}u'(x)\xi(t) \bigg]dx\\
&&\qquad + \ \sum_{n\geq 1}\delta(t-{T}_n)\int_{\R} [\delta(x-x_0)-\rho(x,t^-)]u(x)dx.\nonumber
 \end{eqnarray}
 We have formally set $dW(t)=\xi(t)dt$ with $\xi(t)$ a white noise process such that
 \begin{eqnarray}
 \langle \xi(t)\rangle = 0,\quad  \langle \xi(t)\xi(t')\rangle=\delta(t-t').
 \end{eqnarray}
   Integrating by parts and using the arbitrariness of $u$ leads to the stochastic partial differential equation (SPDE)
\begin{eqnarray}
 \frac{\partial \rho}{\partial t}&=&D\frac{\partial^2 \rho(x,t)}{\partial x^2} -\frac{\partial F(x)\rho(x,t)}{\partial x}   +\sqrt{2D}\frac{\partial \rho(x,t)}{\partial x}\xi(t)\nonumber \\
 &&\quad +\sum_{n\geq 1}\delta(t-{T}_{n})[\delta(x-x_0)-\rho(x,t^-) ].
 \label{SPDE}
\end{eqnarray}

Finally, define the probability density
\begin{equation}
p(x,t)=\left \langle \E \left [\rho(x,t)\right ]\right \rangle\equiv \left \langle \E \left [\delta(x-X(t)) \right ]\right \rangle ,
\label{defpk}
\end{equation}
where $\langle \cdot \rangle$ and $\E[\cdot]$ denote expectations with respect to the white noise process and the Poisson process, respectively. The former can be carried out using the standard property of It\^o calculus, namely, $dW(t)=W(t+dt)-W(t)$ is statistically independent of the current position $X(t)$. An analogous property holds for the Poisson jump process. More specifically, 
\begin{align}
\left \langle \E \left [\rho(x,t^-)dN(t)\right ]\right \rangle= \left \langle \E \left [\rho(x,t^-)\right ]\right \rangle \E \left [ dN(t)\right ].
\end{align}
since $ X(t)$ for all $t<T_n$ only depends on previous jump times. Moreover,
$N(t)-N(\tau)=\int_{\tau}^tdN(s)$ so that $\int_{\tau}^t\E[dN(s)] = \E[N(t)-N(\tau)]=r (t-\tau)$ and, hence,
$\E[dN(t)]=r dt$. We thus obtain the following equation for $p(x,t)$:
\begin{align}
 \frac{\partial p(x,t)}{\partial t}=-\frac{\partial F(x)p(x,t)}{\partial x}+D\frac{\partial^2 p(x,t)}{\partial x^2}+r\delta(x-x_0)-rp(x,t),
\label{CK}
\end{align}
which reduces to equation (\ref{Liou}) when $D=0$ and $F(x)=f(x,\bary)$. The case $D>0$ was originally written down in Refs. \cite{Evans11a,Evans11b}

\setcounter{equation}{0}
\renewcommand{\theequation}{B.\arabic{equation}}
\section*{Appendix B: The cubic formula}
In this appendix we state the general formula for the roots of a cubic \cite{AS} so that we can determine $x_-,\mu,\omega$ as a function of $\bary$ in equation (\ref{ppp}). Consider the cubic
\begin{equation}
Q(x)=x^3+a_2x^2+a_1x+a_0.
\end{equation}
Let
\begin{equation}
u=a_2^2-3a_1,\quad v= 2a_2^3-9a_2a_1+27a_0
\end{equation}
and
\begin{equation}
\label{Cuv}
C=\left (\frac{v+\sqrt{v^2-4u^3}}{2}\right )^{1/3}.
\end{equation}
$C$ can be taken to be any cube root.
The three roots of the cubic are then
\begin{align}
x_k&=-\frac{1}{3}\left (a_2+\xi^k C+\frac{u}{\xi^kC}\right ),\quad k=0,1,2,
\end{align}
where
\begin{equation}
\xi=\frac{-1+\sqrt{-3}}{2}.
\end{equation}
In \S 4 we consider the specific cubic
\begin{equation}
Q(x)=\bary-x(a-x)(x-1)=x^3-(a+1)x^2+ax+\bary,
\end{equation}
so that
\begin{equation}
a_2=-[a+1],\quad a_1=a,\quad a_0=\bary.
\end{equation}
and
\begin{equation}
u=(1+a)^2-3a,\quad v=- 2(1+a)^3+9a(1+a)+27\bary .
\end{equation}


\begin{thebibliography}{10}

\bibitem{AS} {\sc M. Abramowitz and I. Stegun (Eds.)} {\em Handbook of mathematical functions: with formulas, graphs, and mathematical tables}, Dover Books, 2000

\bibitem{Acebron05} {\sc J. A. Acebron, L. L. Bonilla, C. L. Perez Vicente, P. Ritort and R. Spigler,} {\em The Kuramoto model: A simple paradigm for synchronization phenomena.} {Rev. Mod. Phys.} {\bf 77} (2005), pp. 137-185 .

\bibitem{Berez10} {\sc A. M. Berezhkovskii, C. Sample and S. Y. Shvartsman,} {\em How
long does it take to establish a morphogen gradient?}
{Biophys. J.} {99} (2010) pp. L59-L61 



\bibitem{Berez11} {\sc A. M. Berezhkovskii, C. Sample and S. Y. Shvartsman,} {\em Formation
of morphogen gradients: local accumulation time.}
{Phys Rev E} {83} (2011) 051906 

\bibitem{Berez11a} {\sc A. M. Berezhkovskii and S. Y. Shvartsman,}  {\em Physical interpretation of mean local accumulation time of morphogen gradient formation,}
{J. Chem. Phys.} {135} (2011) 154115

\bibitem{Bressloff19} {\sc P. C. Bressloff, S. D. Lawley and P. Murphy,} {\em Protein concentration gradients and switching diffusions,} {Phys. Rev. E} {99} (2019) 032409 



\bibitem{Bressloff20} {\sc P. C. Bressloff,} {\em Occupation time of a run-and-tumble particle with resetting,} {Phys. Rev. E}, {102} (2020) 042135 

\bibitem{Bressloff20a}  {\sc P. C. Bressloff,} {\em Switching diffusions and stochastic resetting,} {J. Phys. A} {53} (2020) 275003.

\bibitem{Bressloff21} {\sc P. C. Bressloff,}  {\em Accumulation time of stochastic processes with resetting,} {J. Phys. A} {54} (2001) 354001.



\bibitem{Bressloff24a}  {\sc P. C. Bressloff,} {\em Global density equations for interacting particle systems with stochastic resetting: from overdamped Brownian motion to phase synchronization,} {Chaos} {34}, (2024) 043101.

\bibitem{Bressloff25}  {\sc P. C. Bressloff,} {\em Kuramoto model with stochastic resetting and coupling through an external medium,} Preprint http://arxiv.org/abs/2411.15534 (2024).



\bibitem{Burkitt06} {\sc A. N. Burkitt,}
{\em A review of the integrate-and-fire neuron model: I. Homogeneous synaptic input,} Biological Cybernetics 95 (2006) pp. 1-19.

\bibitem{Terman}
{\sc G. B. Ermentrout abd D. Terman,}: {\em Mathematical Foundations of Neuroscience.}
\newblock Springer, Berlin, 2010.

  \bibitem{Eule16} {\sc S. Eule S and J. J. Metzger,}  {\em Non-equilibrium steady states of stochastic processes with intermittent resetting,} {New J. Phys. } {18} (2016) 033006
 \bibitem{Evans11a} {\sc M. R. Evans and S. N. Majumdar,} {\em Diffusion with stochastic resetting,} {Phys. Rev. Lett.} {106,} (2011) 160601.

\bibitem{Evans11b} {\sc M. R. Evans and S. N. Majumdar,}  {\em Diffusion with optimal resetting,} {J. Phys. A} {44,} (2011) 435001.

\bibitem{Evans14} {\sc M. R. Evans and S. N. Majumdar,}  {\em Diffusion with resetting in arbitrary spatial dimension,} {J. Phys. A} {47,} (2014) 285001.
 
 \bibitem{Evans18} {\sc M. R. Evans and S. N. Majumdar,}  {\em Run and tumble particle under resetting: a renewal approach,} {J. Phys. A: Math. Theor. Math. Theor.} {51,}  (2018) 475003


\bibitem{Evans20} {\sc M. R. Evans, S. N. Majumdar and G. Scher,}  {\em Stochastic resetting and applications,} {J. Phys. A: Math. Theor.} {53}, (2020) 193001.

\bibitem{Fenichel79} {\sc N. Fenichel,} {\em Geometric singular perturbation theory,} J. Diff. Eq.
31 (1979) pp. 53-91.

\bibitem{Jelbart20} {\sc S. Jelbart, M. Wechselberger,} {\em Two-stroke relaxation oscillations,} Nonlinearity 33 (2020) pp. 2364-2408.

\bibitem{Jones95} {\sc C. K. R. T. Jones,} {\em Geometric Singular Perturbation Theory,} In Dynamical Systems, Lecture Notes in Mathematics, Volume 1609, R. Johnson editor, Springer-Verlag, Berlin (1995) pp. 44-118.

\bibitem{Keener81} {\sc J. P. Keener, F. C. Hoppensteadt and J. Rinzel ,} {\em Integrate-and-fire models of nerve membrane response to oscillatory input,}{SIAM J. Appl. Math.} {41,} (1981) pp. 503-517.

\bibitem{Kuramoto84} {\sc Y. Kuramoto} {\em Chemical Oscillations, Waves, and Turbulence} Springer 1984.

\bibitem{Kus14} {\sc L. Kusmierz, S. N. Majumdar, S. Sabhapandit and G. Schehr,} {\em First order transition for the optimal search time of Levy flights with resetting,} { Phys. Rev. Lett.} {113,} (2014) 220602

\bibitem{Majumdar15}  {\sc S. N. Majumdar, S. Sabhapandit and G. Schehr,} {\em Dynamical transition in the temporal relaxation of stochastic processes under resetting.} {Phys. Rev. E} {91,} (2015) 052131 

\bibitem{Majumder24} {\sc R. Majumder, R. Chattopadhyay and S. Gupta,} {\em Kuramoto model subject to subsystem resetting: How resetting a part of the system may synchronize the whole of it,} {Phys. Rev. E} {109}, (2024) 064137.

\bibitem{Nagar16} {\sc A. Nagar A and S. Gupta,} {\em Diffusion with stochastic resetting at power-law times,} {Phys. Rev. E}
{93} (2016) 060102 (R)

\bibitem{Ott08} {\sc E. Ott and T. M. Antonsen,} {\em Low dimensional behavior of large systems of globally coupled oscillators,} {Chaos} {18} (2008)  037113.

\bibitem{Ozawa24} {\sc A. Ozawa and H. Kori,} {\em Two distinct transitions in a population of coupled oscillators with turnover: desynchronization and stochastic oscillation quenching,} {Phys. Rev. Lett.} {133}, (2024) 047201.

\bibitem{Pal15} {\sc A.Pal,} {\em Diffusion in a potential landscape with stochastic resetting,} {Phys. Rev. E} {91} (2015) 012113  

\bibitem{Pal16} {\sc A. Pal, A. Kundu and M. R. Evans,} {\em Diffusion under time-dependent resetting,} {J. Phys. A: Math. Theor.} {49} (2016) 225001





\bibitem{Sarkar22} {\sc M. Sarkar and S. Gupta,} {\em Synchronization in the Kuramoto model in presence of stochastic resetting,} {Chaos} {32}  (2022) 073109.

\bibitem{Schwab12} {\sc D. J. Schwab, G. G. Plunk and P. Mehta,} {\em Kuramoto model with coupling through an external medium,} {Chaos} {22} (2012) 043139



\bibitem{Strogatz00} {\sc S. H. Strogatz,} {\em From Kuramoto to Crawford: Exploring the onset of synchronization in populations of coupled oscillators,} {Physica D} {143} (2000) pp. 1-20.



\bibitem{Wech20} {\sc M. Wechselberger,} {\em Geometric Singular Perturbation Theory Beyond the Standard Form} Springer, 2020.



\end{thebibliography}
\end{document}